\documentstyle[pra,aps,twocolumn]{revtex}

\begin{document}
\draft
\twocolumn[\hsize\textwidth\columnwidth\hsize\csname @twocolumnfalse\endcsname
\title{\hfill {\small ITP Preprint Number NSF-ITP-95-51}  \\
\vspace{10pt}
Paired States in the Even Integer Quantum Hall Effect}
\author{A. M. Tikofsky and S. A. Kivelson\cite{UCLA}}
\address{Institute for Theoretical Physics, University of California\\
Santa Barbara, CA 93106}

\date{June 12, 1995}
\maketitle

\begin{abstract}
We argue that a new type of quantum Hall state requiring 
non-perturbative Landau level mixing arises at low electron density.
In these states, up and down spin electrons pair
to form spinless bosons that condense into a bosonic quantum Hall state.
We describe a wavefunction for a paired quantum 
Hall state at $\nu=2$ and argue that it is stabilized by a BCS instability
arising in flux attachment calculations.
Based on this state, we derive a new global phase
diagram for the integral quantum Hall effect with spin.
Additional experimental implications are discussed.
\end{abstract}
]

It is common to distinguish two different types of quantum Hall effects:
the integer and the fractional \cite{Girvin}.  When an integer
number of Landau levels are filled and the kinetic energy cost for 
occupying higher Landau levels is large enough,
the occupation of higher Landau levels encouraged by
many-body interactions can be treated perturbatively.
In the fractional quantum Hall effect, many-body interactions
stabilize a ground state with a gap to excitations even though
only a fraction of a Landau level is occupied.
It is assumed that only a single Landau level is partially occupied,
because of the large kinetic energy cost for occupying higher
Landau levels, and that this state is adiabatically connected to the state
with higher Landau level occupation.
We will argue that this traditional picture is incomplete.
When the energy gap between Landau levels is small enough,
there is a new even integer quantum Hall effect that is distinct
from the old one, and whose stability relies on the 
existence of many-particle interactions and higher Landau
level occupation.

This work is motivated by the fact that 
the topology of the experimental phase diagram is inconsistent
with an integral quantum Hall effect insensitive to
many-particle interactions \cite{Jiang1,Jiang2,Pepper}.
Of particular interest is the
existence of a direct second order phase transition from 
$\nu=2$ to $\nu=0$ with the usual quantum Hall critical exponents
(See Fig. 1; $\nu$ is defined as the value of the Hall conductivity
$h\sigma_{xy}/e^2$ in fundamental units).
In the non-interacting picture, the system is required to
pass through an intermediate $\nu=1$ phase in the transition from
$\nu=2$ to $\nu=0$ \cite{Girvin}.
However, there is no experimental evidence for
even a very thin region separating $\nu=2$ and $\nu=0$.
We therefore postulate the existence of a new 
quantum Hall state, which we call the $2b$ state, that is
distinct from the non-interacting spin-unresolved $2a$ state.
The transition from the $2a$ to the $2b$ state
would be first order in the absence of disorder
and is probably second order in the presence of disorder.

The global phase diagram implies that there are direct transitions
between even-integer quantum Hall states in low
magnetic fields.
In fact, at low enough magnetic fields, the odd-integer quantum Hall
effect is known to disappear \cite{Jiang1}.
In addition, the transition directly
from $\nu=0$ to $\nu=2$ and back to $\nu=0$ at low magnetic fields,
shown in the phase diagram, has also been observed experimentally
\cite{Jiang2,Pepper}. 

\input epsf
\begin{figure}
\epsfxsize=8.5 cm
\epsfbox{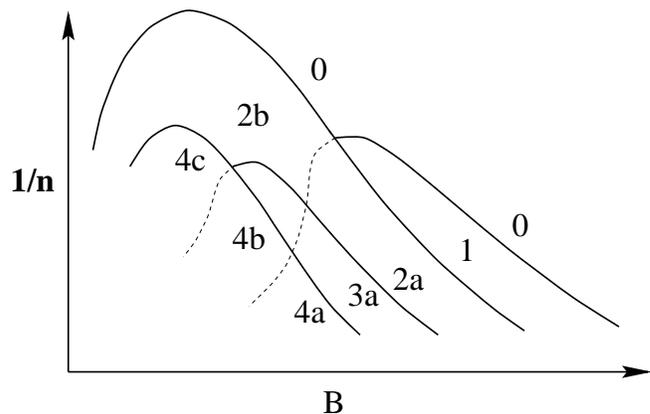}
\caption{The new global phase diagram for the Integral Quantum Hall Effect.
Transitions between the same quantized hall conductance are 
indicated by dashed lines.}
\end{figure}

The essential idea behind the $2b$ quantum Hall state
is that up and down spin electrons pair to form a
spinless boson that condenses into a
bosonic many-body fractional quantum Hall state.
It is clear this state must have significant occupation of
higher Landau levels because the
conventional spin unpolarized $2a$ state is the unique state
at $\nu=2$ in the lowest Landau level.
For purposes of illustration, we describe
a spin-singlet many electron wavefunction \cite{Herbut} 
for the $2b$ state given by
\begin{equation}
\label{wfn}
\Phi=\prod_{k<l,\sigma}(z^\sigma_k-z^\sigma_l)\ {\rm Per} \left[
F(|z_i-z_j|) \right]
{\large e}^{-\sum_i |z_i|^2/4l^2}
\end{equation}
where $z_i$ is the coordinate of the i$^{th}$ electron,
$z^\sigma_k$ is the coordinate of the k$^{th}$ spin $\sigma$
electron, $l$ is the magnetic length, and Per denotes the 
Permanent of the symmetric matrix whose $(i,j)^{th}$
component is $F(|z_i-z_j|)$.
Given an order N matrix M$_{ij}$, the permanent
Per$(M)=H_1(M)$ where 
$H_\alpha(M)=\sum_{P}(\alpha)^{S\{P\}}\prod_{ab}M_{ab}$,
the product is over the $N/2$ pairs ${ab}$, the sum is over the 
possible permutations of these pairs, and $S\{P\}$ is $-1$
or $+1$ if the permutation $P$ is odd or even, respectively
\cite{Wilczek,Read}.
The Pfaffian of a matrix M is Pf$(M)=H_{-1}(M)$.

The behavior of the $2b$ state described in Eq. (\ref{wfn}) is governed
by the function $F$.  For $F=1$, $\Phi$ is the conventional $2a$ state.
On the other hand, if we take $F$ to be a
delta function then we can define a pair coordinate 
and $\Phi$ vanishes as the second power
of the pair coordinate as two pairs approach each other.
This wavefunction is therefore equivalent to a $\nu'=1/2$
Laughlin state for bosonic pairs of charge $e^*=2e$.  Its Hall
conductivity is $\sigma_{xy}=(e^*)^2\nu'/h=2e^2/h$.
If $F$ is short-ranged then it can be thought of as a pair wavefunction
and its effective size is the coherence length.  For distances much
longer than this coherence length,
$F$ acts like a delta-function and $\Phi$
appears to be a $\nu'=1/2$ state.  The $2a$ to $2b$ transition is
thus characterized by the divergence of this coherence length.

At first, the existence of a pairing state
stabilized by repulsive Coulomb interactions seems antintuitive.
Moreover, because $\Phi$ in Eq. (\ref{wfn}) is constructed 
by occupying higher
Landau levels, we lose kinetic energy that must
be compensated by a gain in interaction energy.
However, we only lose interaction energy among the two
electrons in a pair; we gain repulsive energy from pairing
because the $2b$ state acts like a Laughlin $\nu'=1/2$ state for pairs.
As a first approximation, let us ignore the energy cost
in the $2b$ state for
forming tightly bound pairs and calculate the interaction
energy for all electrons not in the same pair.  This is simply
the Coulomb energy of a $\nu'=1/2$ state of spinless
charge $e^*=2e$ bosons.  
Laughlin's interpolation formula
for $E_m$ \cite{Girvin},
the Coulomb energy of a projected $\nu=1/m$ state,
gives an energy per electron for the $2b$ state of
$E^*_{2b}=-.49 (e^*)^2/\epsilon l^*=-1.39 e^2/\epsilon l$.
In contrast, the energy of the $2a$ state is
$E_{2a}=-\sqrt{\pi/8}\ e^2/\epsilon l$.
Forming a paired state has yielded an energy gain per pair of
$2(E_{2a}-E^*_{2b})=1.53 e^2/\epsilon l$ which must more than compensate
the Coulomb and kinetic
energy lost by pairing if the $2b$ state is to be energetically
stable.  This requirement is made easier by the
softening of the Coulomb repulsion at short distances due to
the finite thickness of the electron gas \cite{Girvin,Wilczek}.

As the magnetic field decreases, 
Coulomb energy becomes more important and the Coulomb energy gain of
a $2b$ state with respect to a $2a$ state can more easily
outweigh the loss of kinetic energy.
To get a feeling for the magnitude of the magnetic field at which
such a transition could occur, we study the properties of the
$2a$ state perturbatively in powers of the ratio of the interaction
energy to the kinetic energy,
\begin{equation}
y={e^2/\epsilon l\over  \hbar \omega_c}=l/a^*=\sqrt{B^*\over B}\ ,
\end{equation}
where $a^*=\epsilon \hbar^2 /m^*$ is
the effective Bohr radius and $m^*$ is the band effective mass.
The $2a$ state is exact in the infinite magnetic field or $y=0$ 
limit and is perturbatively stable for sufficiently small $y$.
Failure of perturbation theory, which we take to signal
a transition, would occur at a critical value of $y_c\sim 1$.
For concreteness, we take $y_c=2.3$, the value at which the
second order perturbative expression for the gap of the $2a$ state
vanishes \cite{Skyrmions}.
Given material parameters for
$GaAs-Al_xGa_{1-x}As$ of $a^*=100 \AA$ and $B^*=6.6 T$, this
value of $y_c$ corresponds to a magnetic field of
$B=B^*/y_c^2=1.3T$ which is in the same range as
$.5T$, the experimental magnetic field at which the 
$\nu=0$,$1$, and $2$ quantum Hall states coincide
(see Fig. 1) \cite{Jiang1}.

{\bf BCS Instability:}
We will employ the technique of flux attachment
to show that a spin degenerate electron gas at $\nu=2$ has the
pairing instability necessary to stabilize the $2b$ state \cite{Zhang,dhlee}.
This procedure has the advantage that it implicitly describes
wavefunctions not projected into the lowest Landau level and provides
a motivation, other than energetics, for constructing a paired wavefunction.
We choose as our basis a set of quasiparticles
that are defined as an electron attached to one unit of
spin-relative flux. 
This construction is accomplished by
coupling $j^\sigma_\mu$, the
$\mu^{\rm th}$ component of the current of spin $\sigma$ electrons,
to the gauge field $a^\mu_\sigma$, and requiring that 
\begin{equation}
\label{constraint}
\nabla \times {\bf a}_\sigma({\bf r})=2\pi
\rho_{-\sigma}({\bf r'})\ 
\end{equation}
where $\rho_\sigma({\bf r})$ denotes the density of spin $\sigma$ electrons.
This flux attachment makes unlike spins relative bosons and keeps
like spins relative fermions.
The spin $\sigma$ quasiparticles see a mean magnetic field
\begin{equation}
B_{eff,\sigma}= \nabla \times ({\bf A}-<{\bf a}_\sigma>)\ ,
\end{equation}
which is zero since a $\nu=2$ spin-unpolarized state 
satisfies $\nabla \times {\bf A}=2\pi<\rho_\sigma({\bf r})>$.
At the mean-field level, the quasiparticles form a spin-degenerate
Fermi sea because like spins are non-interacting relative fermions in
zero magnetic field.  We can calculate the effect of fluctuation
corrections on both the system's 
linear-response function and the dressed interparticle
interaction \cite{HLR}.
First of all, we 
define the linear-response function $\Pi$ 
by the statement that a spin-dependent gauge field $A_{ext}^{\sigma'}$
of frequency $\omega$ and wavevector ${\bf q}$ induces a 
current of spin $\sigma$ quasiparticles 
\begin{equation}
J^{\sigma}_\mu(\omega,{\bf q})=\Pi^{\sigma,\sigma'}_{\mu\nu}
(\omega,{\bf q})A^{\sigma'}_{ext,\nu}(\omega,{\bf q}) 
\end{equation}
where $\mu,\nu$ take on the values (0,x,y).
We follow the standard convention of working in Coulomb gauge so that
${\bf q}\cdot{\bf A^\sigma}={\bf q}\cdot{\bf a^\sigma}=0$.
The linear-response kernel of the system
can therefore be described by using only the 4x4 submatrix
$\Pi^{\sigma,\sigma'}_{\alpha\beta}$ in which the subindices $\alpha$
and $\beta$ take on the values
0 and 1 corresponding to the time and transverse component of a vector,
respectively.  In this notation, the response function of the mean-field
degenerate Fermi sea is
\begin{equation}
K^{\sigma,\sigma'}_{\alpha\beta}(q,\omega)
=\delta^{\sigma,\sigma'}\delta_{\alpha\beta}P_{\alpha\alpha}(q,\omega)\ 
\end{equation}
while the bare momentum space interaction is 
\begin{equation}
\label{propagator}
V^{\sigma,\sigma'}_{\alpha\beta}(q,\omega)
={2\pi e^2\over \epsilon q}\delta_{\alpha,o}\delta_{\beta,o}
+{2\pi i\over q}\epsilon_{\alpha\beta}(1-\delta^{\sigma,\sigma'})\ .
\end{equation}
We have assumed a spin-independent
Coulomb interaction between electrons.
In the RPA \cite{Zhang,HLR},
the screened momentum space interaction $D$
and response kernel $\Pi$ satisfy the matrix equations
\begin{equation}
\label{rpa}
D = V+V K D\ ,\ {\rm and}\qquad \Pi = K+K V \Pi\ .
\end{equation}

We now use this flux attachment formalism to show that the
system has an instability towards singlet BCS pairing.
Because we are interested in
spin-singlet pairing and have 
transmuted the particles into relative bosons, the pairing must be
odd parity. 
The BCS instability can be shown to be strongest in
the p-wave channel, the lowest allowed
angular momentum channel.
The screened interaction $D$, given in Eq. (\ref{rpa}), yields the
BCS pair interaction in the singlet channel 
\begin{eqnarray}
V({\bf k},{\bf k'},\omega)&=&D^{\uparrow\downarrow}_{00}(q,\omega)+
({\bf k} \times {\bf k'}) D^{\uparrow\downarrow}_{01}(q,\omega)/q\nonumber\\
&+&|{\bf k} \times {\bf k'}|^2 D^{\uparrow\downarrow}_{11}(q,\omega)/q^2\ .
\end{eqnarray}
where $q=|{\bf k}-{\bf k'}|$ \cite{Wilczek,Bonesteel}.
We assume a gap function with odd $m$-wave pairing,
$\Delta({\bf k})=|\Delta(k)|e^{-im\phi_{\bf k}}$.
Because $D^{\sigma,\sigma'}_{01}$ is pure imaginary and
$V({\bf k},{\bf k'},\omega)$ thus has a non-vanishing imaginary component,
the pairing instability in the BCS gap equation, for a fixed $|m|$,
is strongest for $m<0$.  However, the dominant attraction in all
angular momentum channels comes from
$D^{\uparrow\downarrow}_{11}(q,\omega)$ \cite{Bonesteel,unpublished}.

A spin-singlet wavefunction corresponding to the flux
attachment calculations is given by 
\begin{equation}
\label{wfn2}
\Phi'=\prod_{k,l}{1\over(z^\uparrow_k-z^\downarrow_l)^{\bf *}}
\ {\rm Pf} \left[
(z_i-z_j)G_{ij} \right]
{\large e}^{-\sum_i |z_i|^2/4l^2}
\end{equation}
instead of Eq. (\ref{wfn})
where $G_{ij}=G(|z_i-z_j|)$ and the Pfaffian is defined
in conjunction with Eq. (\ref{wfn}).  The prefactor in Eq. (\ref{wfn2})
implements the
singular gauge transformation described in Eq. (\ref{constraint})
and can be shown to make $\Phi'$ into a spin-singlet wavefunction.
(If we had chosen to implement the gauge transformation via the prefactor
$\prod_{k,l}(z^\uparrow_k-z^\downarrow_l)$ then $\Phi'$ would not be 
a spin-singlet.)
In contrast to the gauge theory description of $\Phi'$,
a similar description of $\Phi$, given in Eq. (\ref{wfn}), would
transmute the electrons into bosonic quasiparticles.
Therefore, a pairing mechanism describing $\Phi$ would 
have to bind two charge $e$ bosons into a charge $2e$ boson \cite{Rice}.
Nevertheless, we believe that both $\Phi$ in Eq. (\ref{wfn})
and $\Phi'$ describe the same physics in the thermodynamic limit.

The physical behavior of the paired state is most easily expressed in
terms of the gauge field and current combinations \cite{Zee}:
\begin{equation}
a^\pm={1\over2}(a^\uparrow\pm a^\downarrow)\ ,\qquad
j^\pm_\mu
=j^\uparrow_\mu\pm j^\downarrow_\mu\ .
\end{equation}
The $j^+$ current only couples to the $a^+$ gauge field
while the $j^-$ current only couples to the $a^-$ gauge field.
A singlet paired state has a gap $\Delta_s$
for single particle excitations in both  
the $j^+$ and $j^-$ channels.  In addition,
the $j^+$ channel has a gapless Goldstone mode whose spectrum is given by the
poles of $a^+$'s propagator.
While fluctuations of the statistical gauge field $a$
do not affect the low-energy spectrum of
the $j^-$ channel \cite{Hall-Insulator}, they 
gap the Goldstone mode of the $j^+$ channel and
give it a quantized $\nu=2$ Hall effect \cite{Zhang,dhlee,HLR}.
Both of these effects are apparent in the screened response functions
$\Pi^\pm=2(\Pi^{\uparrow\uparrow}\pm\Pi^{\uparrow\downarrow})$
where $\Pi$ is defined in Eq. (\ref{rpa}) and spin-symmetry assumed.

{\bf Implications:}
The $2b$ state has a number of different quasiparticle excitations.
Because the paired state describes spinless bosons
at filling fraction $\nu'=1/2$, we can construct the usual quantum Hall
quasiparticles
with charge $\pm e^*\nu'=\pm e$ and statistics $\pm\pi\nu'=\pm\pi/2$.
We refer to these charge 1 spinless semions as holons.
When we bind a single electron to a holon, we make a
charge 0 spin 1/2 semion which we refer to as a spinon. 
In addition, electrons are also quasiparticle excitations and
may be viewed as a bound spinon-holon pair.

Let us now construct quantum Hall hierarchies by using the $2b$ paired state 
as a parent state.  
Because electrons are 
well-defined quasiparticles, one can imagine constructing 
conventional integral quantum Hall states of up and down spin electrons
on top of the parent $2b$ state.  These are called $nb$ states in the
figure.  On the other hand, we can pair up and down spin electrons into
a new boson and form a $\nu'=1/2$ bosonic quantum Hall states of these
new pairs.  The $4c$ state shown in Fig. 1 is an example of such a state.
We can construct yet another hierarchy of paired spin-unpolarized states
based on the condensation of
holons at filling fractions $\nu=4\nu'=8p/(4p+1)$ where $p$ is
an integer \cite{Girvin}. 

In analogy to the $2b$ state, we have constructed other singlet paired
states at filling fraction $\nu=2/(2p+1)$ where p is an integer.
These states correspond
to bosonic quantum Hall states of charge $2e$ bosons at filling
fraction $\nu'=1/2(2p+1)$.
If we try to construct triplet paired states, spin-rotation
symmetry requires them to be spin-polarized.
However, there is no BCS instability for spin-polarized states; in
fact, the $\omega=0$ BCS interaction is repulsive in all
angular momentum channels \cite{Bonesteel}.
Therefore, these states are unlikely to
exist in nature unless spin symmetry is broken.
In fact, there is evidence
for triplet pairing in bilayer systems which do not have
pseudospin-rotation (layer index) symmetry \cite{bilayer}.

Let us now address the nature of the transitions
between the various states shown in Fig. 1, in the
presence of dirt.  It is well established
that the transitions between daughter and parent quantum Hall states
are second order and are all similar to each other.
These transitions can be viewed as the quantum percolation transition
of a single edge state with the selection rule \cite{Kivelson}
that only one state can undergo this transition at a time.
In this language, the $2a$ state has 2 edge modes,
one composed of each spin polarization, and the
$2a$ to $1$ transition is associated with the percolation of
the down spin edge state.
Alternatively, for small Zeeman energy, the $2a$ state 
can be viewed as having a charged holon edge mode and a neutral spinon
edge mode.  The transition from $2a$ to $2b$ can therefore be viewed
as spinon percolation.  We postulate that this transition is similar to
the $2a$ to $1$ transition but that the
critical behavior is visible in the spin degrees of freedom and that
the system is a spin-metal at the critical point.
In addition, we identify the tetracritical
point in Fig. 1 as the intersection of the phase boundaries
associated with the two order parameters:
the Girvin-MacDonald order parameter defined by attaching one unit
of ${\bf a}^\uparrow$ flux to the up spin electron creation operator
is nonzero in both the $1$ and $2a$ phases;
the order parameter defined by attaching one unit of
$({\bf a^\uparrow}+{\bf a^\downarrow})/2$ flux
to the charge $2e$ pair creation operator is nonzero in both the $2a$
and $2b$ phases \cite{Girvin}.

{\bf Other Experimental Implications:}
It has been conjectured that there is a universal value for the
resistance at quantum hall phase transitions \cite{Kivelson}.
Recent experiments have in fact found universal values for the
resistance at quantum hall liquid
to insulator transitions \cite{Jiang1,Jiang2,Pepper,Tsui}.
According to the pairing theory,
the transition between $\nu=2$ and $\nu=0$ corresponds to a
$\nu'=1/m=1/2$ quantum hall liquid to insulator transition
of charge $e^*=2e$ quasiparticle with the corresponding {\it
universal resistances} \cite{Kivelson}:  $\rho_{xx}=h/(e^*)^2=h/4e^2$
and $\rho_{xy}=mh/(e^*)^2=h/2e^2$.  This prediction is consistent with
the observed values for the resistances in the experiments of Wong
{\it et. al.} \cite{Jiang1}.
(It is important to note that there are experiments at higher magnetic
fields that find different resistances at the $\nu=2$ to insulator
phase transition or a different critical exponent \cite{Pepper}.
This discrepancy is not yet understood.)

Our theory predicts a spin gap $\Delta_s$ associated with pairing.
In the presence of dirt, this may become a pseudogap.
Nonetheless, this gap
should be experimentally observable by studying the density of
free spins as a function of $\delta B=B-\pi\rho$ \cite{Skyrmion-exp}.
The density of free spins should grow linearly with $|\delta B|$
in the $2a$ state while it should be strongly suppressed in the $2b$ state
for temperatures below $\Delta_s$, even for $\delta B\ne 0$.
In addition, the Zeeman splitting should act as a pair breaker and
decrease the spin gap $\Delta_s$.
We therefore expect the $2b$ state to be destroyed when
the Zeeman energy exceeds the spin gap $\Delta_s$.
Indeed, the triple point, where $\nu=0$,$1$, and $2$ coexist in Fig. 1,
is pushed to lower fields as the Zeeman energy is increased by tilting
the magnetic field \cite{Jiang1},

The point contact conductance of the $2b$ state will be very
different than that of a $2a$ state.
The $2a$ state has 2 Fermi-liquid edge modes with nonzero conductance
at zero temperature \cite{Wen}.  In contrast, the $2b$ state has only a single
$\nu'=1/2$ edge mode of charge $e^*$ carriers.
For sufficiently low temperatures, the point contact tunneling
will be non-Fermi-liquid like and the conductance will
vanish as $T^2$ at very low 
temperatures \cite{Wen,qhe-tunneling}.
In addition, the scaling function for the resonance line shape,
in the limit of small temperature, is known exactly for $\nu'=1/2$.
In the limit of strong backscattering (large gate voltage),
the conductance is dominated by single electron
tunneling and hence is thermally activated because of the gap
$\Delta_s$ to single particle excitations.  This should provide
an experimental measure of $\Delta_s$.

In future work, we will explore in detail the energetic properties of
the proposed paired wavefunctions for finite-size systems.

We are grateful to H. W. Jiang for sharing his results with us
prior to publication.  We have also benefited from many
useful conversations with N. E. Bonesteel, A. H. Castro-Neto,
M.P.A. Fisher, D. Morse, S. L. Sondhi, and A. Zee.
This work has been supported by the National Science Foundation
under grant PHY94-07194 at the Institute for Theoretical Physics
and grant DMR93-12606 at UCLA.

\end{document}